\documentclass[epj]{svjour} 
\usepackage{graphics} 
\usepackage{amsmath} 
\begin{document} 
\title{Facets of confinement and dynamical chiral symmetry breaking
}
\author{P.\ Maris\inst{1} \and A.\ Raya\inst{2}%
\and C.D.\ Roberts\inst{3,}\inst{4} \and S.M.\ Schmidt\inst{5} 
}                     
%
%
\institute{Department of Physics, North Carolina State University, Raleigh NC 
27695-8202, USA \and %
Instituto de F\'{\i}sica y Matem\'aticas, Universidad Michoacana de San 
Nicol\'as de Hidalgo,\\ Apartado Postal 2-82, Morelia, Michoac\'an, M\'exico 
\and 
Physics Division, Argonne National Laboratory, Argonne IL 60439-4843, USA \and %
Fachbereich Physik, Universit\"at Rostock, D-18051 Rostock, Germany \and%
Helmholtz-Gemeinschaft, Ahrstrasse 45, D-53175 Bonn, Germany 
} 
\date{\vspace*{-1ex}}
%
\abstract{The gap equation is a cornerstone in understanding dynamical chiral
symmetry breaking and may also provide clues to confinement. A
symmetry-preserving truncation of its kernel enables proofs of important
results and the development of an efficacious phenomenology.  We describe a
model of the kernel that yields: a momentum-dependent dressed-quark
propagator in fair agreement with quenched lattice-QCD results; and chiral
limit values: $f_\pi^0= 68\,$MeV and $\langle \bar q q \rangle = -(190\,{\rm
MeV})^3$. It is compared with models inferred from studies of the gauge
sector.
} 
\maketitle 
 
\section{Introduction} 
\label{intro} We will address these topics from the perspective of QCD's 
Dyson-Schwinger equations (DSEs) \cite{bastirev}.  The DSEs are a keystone in
proving renormalisability and provide a generating tool for perturbation
theory.  The latter point is very important in applications to low-energy
phenomena because it means that the model-dependence which necessarily
appears in continuum studies of nonperturbative QCD is restricted to the
infrared; i.e, to momentum-scales $\stackrel{<}{\mbox{\tiny$\sim$}} 1\,$
GeV$^2$.  This feature has successfully been exploited in applications to the
spectra \cite{pmspectra1,pmspectra2}, and strong \cite{pmpipi} and
electroweak \cite{pmew} interactions of mesons.  It also mitigates, to a
useful extent, some of the problems with the approach; e.g., it helps in
developing reliable truncations for the coupled system of DSEs.
 
Recent successes are founded on an accurate understanding of dynamical chiral
symmetry breaking (DCSB), and its role in resolving the dichotomy of the pion
\cite{mrt98} (as both a Goldstone mode and a massless bound state of massive
constituents) and its relation to the long-range behaviour of the effective
coupling between quarks \cite{hmr98}.
 
\section{Dynamical chiral symmetry breaking} 
\label{sec:2} This is a purely nonperturbative phenomenon that can be studied 
via QCD's gap equation: 
\begin{eqnarray} 
\nonumber 
\lefteqn{S(p)^{-1}  =  \,(i\gamma\cdot p + m)}\\ 
&&  +\, \int\frac{d^4q}{(2\pi)^4} \, g^2 D_{\mu\nu}(p-q) 
\frac{\lambda^a}{2}\gamma_\mu  S(q) \Gamma^a_\nu(q,p) \,, \label{gendse} 
\end{eqnarray} 
wherein: $m$ is the current-quark bare mass; $g$ is the coupling constant; 
$D_{\mu\nu}(p-q)$ is the dressed-gluon propagator; $\Gamma^a_\nu(q,p)$ is the 
dressed-quark-gluon vertex; and the solution is the dressed-quark propagator 
\begin{equation} 
S(p)=  \frac{1}{i \gamma\cdot p \, A(p^2) + B(p^2) }
= \frac{Z(p^2)}{i \gamma\cdot p \,  + M(p^2) }\,. \label{sinvp} 
\end{equation} 
(Equation (\ref{gendse}) is the unrenormalised equation: renormalisation will 
only be mentioned as necessary.) 
 
As noted in the Introduction, one can use the gap equation to evaluate the 
fermion self-energy perturbatively and thereby obtain 
\begin{equation} 
B(p^2) = m \left( 1 - \frac{\alpha}{\pi} \ln\left[\frac{p^2}{m^2}\right] + 
\ldots \right)\,, 
\end{equation} 
wherein the ellipsis denotes two- and higher-loop contributions.  It is 
apparent that each term is proportional to the current-quark mass and vanishes 
as $m\to 0$.  Hence, in perturbation theory, if there is no quark mass to begin 
with, interactions do not generate one; i.e., DCSB is impossible in 
perturbation theory. 
 
In the chiral limit, namely, $m=0$, Eq.~(\ref{gendse}) always admits the
trivial solution $B(p^2)\equiv 0$; i.e., the solution accessible
perturbatively.  However, as is characteristic of gap equations, when the
integrated strength of the kernel exceeds some critical value, a $B(p^2) \neq
0$ solution is generated.  This nonperturbative, dynamical generation of a
quark mass \textit{from nothing} is DCSB.  The integrated strength of the
kernel can be characterised by an interaction tension, $\sigma^\Delta$, for
which the critical value is \cite{cdresi} $\sigma_c^\Delta \sim 2.5\,$GeV/fm:
this strength provides for DCSB \textit{but only just}.  An acceptable
description of hadrons requires \cite{pmspectra1} $\sigma^\Delta \sim
25\,$GeV/fm, which is an order of magnitude larger.
 
\subsection{Symmetry preserving truncation} 
It is apparent that the gap equation's kernel is formed from a product of the
dressed-gluon propagator and dres\-sed-quark-gluon vertex. It may be
calculated in perturbation theory but that is inadequate for the study of
intrinsically nonperturbative phenomena.  Consequently, to make sensible
statements about DCSB, one must employ an alternative systematic and chiral
symmetry preserving truncation scheme.
 
One must also focus on more than just the gap equation's kernel.  Chiral
symmetry is expressed via the axial-vector Ward-Takahashi identity:
\begin{eqnarray} 
P_\mu \, \Gamma_{5\mu}(k;P) & = & S^{-1}(k_+)\, i\gamma_5 + i\gamma_5 
\,S^{-1}(k_-)\,, \label{avwti} 
\end{eqnarray} 
$k_\pm = k\pm P/2$, wherein $\Gamma_{5\mu}(k;P)$ is the dressed axial-vector 
vertex.  This three-point function satisfies an inhomogeneous Bethe-Salpeter 
equation: 
\begin{eqnarray} 
\nonumber\lefteqn{\left[\Gamma_{5\mu}(k;P)\right]_{tu}= 
\left[\gamma_5\gamma_\mu\right]_{tu} }\\ 
 &  & + \int\frac{d^4 q}{(2\pi)^4} [S(q_+) 
\Gamma_{5\mu}(q;P) S(q_-)]_{sr} K_{tu}^{rs}(q,k;P)\,, 
\end{eqnarray} 
in which $K(q,k;P)$ is the fully-amputated quark-antiquark scattering kernel. 
The Ward-Takahashi identity, Eq.~(\ref{avwti}), means that the kernel in the 
gap equation and that in the Bethe-Salpeter equation are intimately related. 
Therefore a qualitatively reliable understanding of chiral symmetry and its 
dynamical breaking can only be obtained using a truncation scheme that 
preserves this relation, and hence guarantees Eq.~(\ref{avwti}) without a 
\textit{fine-tuning} of model-dependent parameters. 
 
One such scheme exists \cite{truncscheme}.  Its leading-order term is the 
renormalisation-group-improved rainbow-ladder truncation and the general 
procedure provides a means to identify, \textit{a priori}, those channels in 
which that truncation is likely to be accurate. This scheme underlies the 
successful application of a rainbow-ladder model to flavour-nonsinglet 
pseudoscalar mesons and vector mesons \cite{pmspectra1,pmspectra2,pmpipi,pmew}. 
The systematic nature of the scheme has also made possible a proof of 
Goldstone's theorem in QCD \cite{mrt98} and the derivation of a mass formula 
that unifies the light- and heavy-quark sectors \cite{cdrlc01}. 
 
\begin{figure}[t] 
\centerline{\resizebox{0.50\textwidth}{!}{\includegraphics{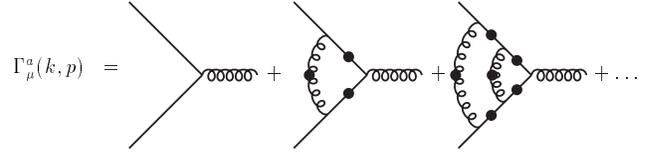}}} 
\caption{\label{fig1} Planar dressed-quark-gluon vertex obtained by neglecting 
contributions associated with explicit gluon self-interactions. Solid circles 
indicate fully dressed propagators.  The quark-gluon vertices are not dressed.} 
\end{figure} 
 
In this scheme, as in perturbation theory, it is impossible, in general, to
obtain complete closed-form expressions for the kernels of the gap and
Bethe-Salpeter equations.  However, for the planar dressed-quark-gluon vertex
depicted in Fig. \ref{fig1}, closed forms can be obtained and a number of
significant features illustrated \cite{will} when one uses the following
model for the dressed-gluon line \cite{mn83}
\begin{equation} 
\label{mnmodel} g^2 \, D_{\mu\nu}(k) = 
\left[\delta_{\mu\nu} - \frac{k_\mu k_\nu}{k^2}\right] (2\pi)^4\, {\cal G}^2
\, \delta^4(k)\,,
\end{equation} 
where ${\cal G}$, measured in GeV, sets the model's mass-scale.  This form is a 
precursor to the class of renormalisation-group-improved models.  It has many 
positive features in common with that class and, furthermore, its particular 
momentum-dependence works to advantage in reducing integral equations to 
character-preserving algebraic equations.  Naturally, there is a drawback: the 
simple momentum dependence also leads to some model-dependent artefacts, but 
they are easily identified and hence not cause for concern. 
 
It is a general result \cite{will} that with any vertex whose diagrammatic 
content is known explicitly; e.g., Fig. \ref{fig1}, it is always possible to 
construct a unique Bethe-Salpeter kernel which ensures the Ward-Takahashi 
identities are automatically fulfilled: that kernel is necessarily non-planar. 
This becomes transparent with the model in Eq.\ (\ref{mnmodel}), using which 
the gap equation obtained with the vertex depicted in Fig.\ \ref{fig1} reduces 
to an algebraic equation, irrespective of the number of dressed-gluon rungs 
that are retained, and the same is true of the Bethe-Salpeter equations in 
every channel: pseudoscalar, vector, etc. 
 
\begin{table}[b] 
\centerline{\rule{0.48\textwidth}{0.1ex}} \caption{\label{tab1} $\pi$ and 
$\rho$ meson masses obtained with ${\cal G}= 0.48\,{\rm GeV}$.   (Dimensioned 
quantities in GeV.)  $n$ is the number of dressed-gluon rungs retained in the 
planar vertex, see Fig.~\protect\ref{fig1}, and hence the order of the 
vertex-consistent Bethe-Salpeter kernel.} 
\begin{tabular*} 
{\hsize}{l@{\extracolsep{0ptplus1fil}} 
|c@{\extracolsep{0ptplus1fil}}c@{\extracolsep{0ptplus1fil}} 
c@{\extracolsep{0ptplus1fil}}c@{\extracolsep{0ptplus1fil}}} 
\hline
%
 & $M_H^{n=0}$ & $M_H^{n=1}$ & $M_H^{n=2}$ & $M_H^{n=\infty}$\\\hline 
$\pi$, $m=0$ & 0 & 0 & 0 & 0\\ 
$\pi$, $m=0.011$ & 0.152 & 0.152 & 0.152 & 0.152\\\hline 
$\rho$, $m=0$ & 0.678 & 0.745 & 0.754 & 0.754\\ 
$\rho$, $m=0.011$ & 0.695 & 0.762 & 0.770 & 0.770 \\ 
%
\hline 
\end{tabular*} 
\end{table} 
 
Results for the $\pi$ and $\rho$ are illustrated in Table \ref{tab1}.  It is 
evident that, irrespective of the order of the truncation; i.e., the value of 
$n$, the number of dressed gluon rungs in the quark-gluon vertex, the pion is 
massless in the chiral limit.  (NB.\ This pion is composed of heavy 
dressed-quarks, as is evident in the calculated scale of the dynamically 
generated dressed-quark mass function: $M(0) \approx 0.5\,$GeV.) The 
masslessness of the $\pi$ is a model-independent consequence of the consistency 
between the Bethe-Salpeter kernel and the kernel in the gap equation. 
Furthermore, the bulk of the $\rho$-$\pi$ mass splitting is present in the 
chiral limit and with the simplest ($n=0$; i.e., rainbow-ladder) kernel, which 
makes plain that this mass difference is driven by the DCSB mechanism: it is 
not the result of a finely adjusted hyperfine interaction.  Finally, the 
quantitative effect of improving on the rainbow-ladder truncation; i.e., 
including more dressed-gluon rungs in the gap equation's kernel and 
consistently improving the kernel in the Bethe-Salpeter kernel, is a 10\% 
correction to the vector meson mass.  Simply including the first correction 
($n=1$; i.e., retaining the first two diagrams in Fig.\ \ref{fig1}) yields a 
vector meson mass that differs from the fully resummed result by 
$\stackrel{<}{\mbox{\tiny$\sim$}}1$\%.  The rainbow-ladder truncation is 
clearly accurate in these channels. 
 
\subsection{Rainbow-ladder truncation} 
\label{sec:rl} The renormalisation-group-improved rainbow-ladder truncation is 
based on the fact that in QCD, on the domain for which $Q^2 := (k-q)^2 \sim 
k^2\sim q^2$ is large and spacelike, 
\begin{eqnarray} 
\nonumber \lefteqn{M(q,k;P) := g^2(\mu^2)\, D_{\mu\nu}(k-q) \,} \\ 
&& \nonumber \times 
\left[\rule{0mm}{0.7\baselineskip} \Gamma^a_\mu(k_+,q_+)S\,(q_+) \right] 
\otimes 
\left[ \rule{0mm}{0.7\baselineskip}S(q_-)\,\Gamma^a_\nu(q_-,k_-) \right] \\ 
&= & \nonumber 
4\pi\,\alpha(Q^2)\, D_{\mu\nu}^{\rm free}(k-q)\,\\ 
&& \times 
\left[\rule{0mm}{0.7\baselineskip} 
        \frac{\lambda^a}{2}\gamma_\mu \,S^{\rm free}(q_+)\right]\otimes 
\left[\rule{0mm}{0.7\baselineskip}S^{\rm free}(q_-)\, 
        \frac{\lambda^a}{2}\gamma_\nu\right]\,, \label{dressedL} 
\end{eqnarray} 
where $\alpha(Q^2)$ is the strong running coupling and, e.g., $S^{\rm free}$ is 
the free quark propagator.  The dressed-ladder truncation supposes that 
Eq.~(\ref{dressedL}) is valid for all momenta and is thus an assumption about 
the long-range ($Q^2 \stackrel{<}{\mbox{\tiny$\sim$}} 1\,$ GeV$^2$) behaviour 
of the interaction.  Requiring that Ward-Ta\-ka\-ha\-shi identities be 
automatically fulfilled by any truncation entails \cite{truncscheme,will} that 
the kernel in the renormalised gap equation is expressed through 
\begin{equation} 
 {\cal G}(Q^2) \,D_{\mu\nu}^{\rm free}(Q) 
 \frac{\lambda^a}{2}\gamma_\mu S(q) 
\frac{\lambda^a}{2}\gamma_\nu\,, \label{rainbow} 
\end{equation} 
where we have introduced ${\cal G}(Q^2)$, an effective interaction, to 
emphasise that, while ${\cal G}(Q^2)=4\pi \alpha(Q^2)$ is well-de\-ter\-mined 
for $Q^2 \stackrel{>}{\mbox{\tiny$\sim$}} 1\,$ GeV$^2$, the behaviour at $Q^2< 
1\,$GeV$^2$; i.e., at infrared length-scales ($\stackrel{>}{\mbox{\tiny$\sim$}} 
0.2\,$fm), of the kernels in the gap and Bethe-Salpeter equations is unknown. 
 
Contemporary DSE studies employ a model for the infrared behaviour.  The most 
extensively applied is \cite{pmspectra2}: 
\begin{eqnarray} 
\nonumber 
\lefteqn{ \frac{{\cal G}(Q^2)}{Q^2} = \frac{4\pi^2}{\omega^6} D\, Q^2 
{\rm e}^{-Q^2/\omega^2} }\\ 
&& + \, 8\pi\,\frac{ \pi\, \gamma_m } { \ln\left[\tau + \left(1 + 
Q^2/\Lambda_{\rm QCD}^2\right)^2\right]} \, {\cal F}(Q^2) \,, \label{gk2} 
\end{eqnarray} 
where: ${\cal F}(Q^2)= [1 - \exp(-Q^2/[4 m_t^2])]/Q^2$, $m_t$ $=$ $0.5\,$GeV;
$\tau={\rm e}^2-1$; $\gamma_m = 12/(33-2 N_f)$, $N_f=4$; and $\Lambda_{\rm
QCD} = \Lambda^{(4)}_{\overline{\rm MS}}=0.234\,$GeV \cite{pdgold} (NB.\ Eq.\
(\ref{gk2}) gives $\alpha(m_Z^2)= 0.126$. Comparison with a modern value
\cite{pdg}: $0.113^{+0.009}_{ - 0.013}$, means that a smaller $\Lambda_{\rm
QCD}$ is acceptable in the model, if one wants to avoid overestimating the
coupling in the ultraviolet, but not a larger value.)  The true parameters in
Eq.\ (\ref{gk2}) are $D$ and $\omega$, which together determine the
interaction tension in the model.  However, they are not independent: in
fitting, a change in one is compensated by altering the other; e.g., on the
domain $\omega\in[0.3,0.5]\,$GeV, the fitted observables are approximately
constant along the trajectory $\omega \,D = (0.72\,{\rm GeV})^3$.  This
correlation: a reduction in $D$ compensating for an increase in $\omega$,
acts to keep a fixed value of the interaction tension; i.e., the
interaction's integrated strength on the infrared domain.  With
\begin{equation} 
\label{Domega} 
\begin{array}{cc} 
D=(0.96\,{\rm GeV})^2\,,\; & \omega=0.4\,{\rm GeV} 
\end{array} 
\end{equation} 
and current-quark masses 
\begin{equation} 
\label{mqs} 
\begin{array}{cc} 
m_u^{1\,{\rm GeV}}=5.5\, {\rm MeV} \,,\;&  m_s^{1\,{\rm GeV}}=125\, {\rm MeV}\,, 
\end{array} 
\end{equation} 
one obtains the excellent description of meson observables described in the 
Introduction, which can be illustrated via the calculated values of the 
leptonic decay constant and vacuum quark condensate in Table~\ref{tab2}. 
 
\begin{table}[b] 
\centerline{\rule{0.48\textwidth}{0.1ex}} \caption{Calculated pion decay
constant and vacuum quark condensate.  Experimentally, $f_\pi = 0.092$, and a
best phenomenological value of the condensate is : $-\langle\bar q
q\rangle^0_{1 \,{\rm GeV}}= (0.236 \pm 0.008)^3$ \protect\cite{derek}.  (All
quantities are listed in GeV.)\label{tab2}}
\begin{tabular*} 
{\hsize}{l@{\extracolsep{0ptplus1fil}} 
|c@{\extracolsep{0ptplus1fil}} 
c@{\extracolsep{0ptplus1fil}}} 
\hline
%
 Model & $f_\pi$ & $(-\langle\bar q q\rangle^0_{1 \,{\rm 
 GeV}})^{1/3}$\\ \hline 
 Eqs.\ (\protect\ref{gk2}), (\protect\ref{Domega})~ & 0.092 & 0.24 \\ 
 Eqs.\ (\protect\ref{alkoferraya}), (\protect\ref{arparams})~ & 0.053 & 0.15\\ 
 Eqs.\ (\protect\ref{alkoferraya}), (\protect\ref{rainbow2})~ & 0.099 & 0.25 
%
\\\hline 
\end{tabular*} 
\end{table} 
 
\subsection{Comparison with lattice-QCD simulations} 
\begin{figure}[t] 
\centerline{\resizebox{0.48\textwidth}{!}{\includegraphics{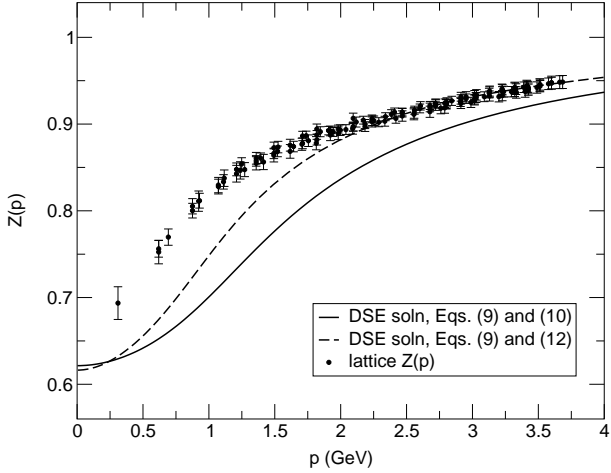}}}
\caption{\label{figZ} Wave function renormalisation.  Solid line: solution of
the renormalised gap equation using Eqs.\ (\protect\ref{gk2}),
(\protect\ref{Domega}); data: lattice simulations
\protect\cite{latticequark}, $m=0.036/a\sim 60\,$MeV; dashed-line: gap
equation solution using Eqs.\ (\protect\ref{gk2}),
(\protect\ref{DomegaL}). (In the DSE studies, the renormalisation point
$\zeta=19\,$GeV and the current-quark mass is $0.6\,m_s^{1\,{\rm GeV}}$ [Eq.\
(\protect\ref{mqs})].)}
\end{figure}

\begin{figure}[t] 
\centerline{\resizebox{0.48\textwidth}{!}{\includegraphics{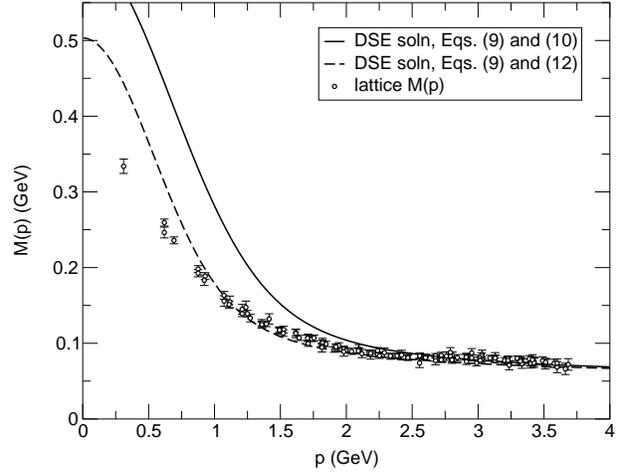}}} 
\caption{\label{figM} Mass function.  Solid line: solution of the
renormalised gap equation using using Eqs.\ (\protect\ref{gk2}),
(\protect\ref{Domega}); data: lattice simulations
\protect\cite{latticequark}, $m=0.036/a\sim 60\,$MeV; dashed-line: gap
equation solution using Eqs.\ (\protect\ref{gk2}), (\protect\ref{DomegaL})
(In the DSE studies, the renormalisation point $\zeta=19\,$GeV and the
current-quark mass is $0.6\,m_s^{1\,{\rm GeV}}$ [Eq.\ (\protect\ref{mqs})].)}
\end{figure}

The scalar functions characterising the renormalised dres\-sed-quark 
propagator: the wave function renormalisation, $Z(p^2)$, and mass function, 
$M(p^2)$, obtained by solving the renormalised gap equation using Eq.\ 
(\ref{gk2}), are depicted in Figs.\ \ref{figZ}, \ref{figM}.  The infrared 
suppression of $Z(p^2)$ and enhancement of $M(p^2)$ are long-standing 
predictions of DSE studies, as Ref.\ \cite{cdragw} makes plain and could be 
anticipated from Ref.\ \cite{bjw}.  This prediction has recently been 
confirmed in numerical simulations of quenched lattice-QCD, as is evident in 
the figures. 
 
It is not yet possible to reliably determine the chiral limit behaviour of
lattice Schwinger functions.  Hence one does not have a lattice estimate for
the quantities in Table \ref{tab2}.  To obtain such an estimate, we used the
DSE model described previously, varying $(D,\omega)$ in order to reproduce
the lattice data.  Our current best result, obtained with
\begin{equation} 
\label{DomegaL} 
\begin{array}{cc} 
D=(0.74\,{\rm GeV})^2\,,\; & \omega=0.3\,{\rm GeV}\,, 
\end{array} 
\end{equation} 
at a current-quark mass of $0.6\,m_s^{1\,{\rm GeV}}\!$, Eq.\
(\protect\ref{mqs}), is also depicted in Figs.\ \ref{figZ}, \ref{figM}, and
yields $f_\pi= 0.094 \,$GeV, $m_\pi = 0.48\,$GeV.  (NB.\ These are values for
a pion-like bound state constructed from constituents whose current-quark
mass is $m \approx 14\,m_u$.  Our DSE-calculated meson-mass versus
current-quark-mass trajectory is consistent with results of recent quenched
lattice-QCD simulations obtained on $m\in [1,2] \,m_s$
\protect\cite{cdrlc01}.) The parameters in Eq.~(\ref{DomegaL}) give chiral
limit results:
\begin{equation}
f_\pi^0= 0.068\,{\rm GeV}\,,\; 
-\langle\bar q q\rangle^0_{1 \,{\rm GeV}} = (0.19\,{\rm GeV})^3\,.
\end{equation}
 
\subsection{Gap equation and QCD's gauge sector} 
Contemporary Landau gauge DSE studies of QCD's gau\-ge sector have been used to 
infer a form of the effective interaction \cite{alkofer}, which is well 
represented by $(x=Q^2/\Lambda_{\rm QCD}^2)$ 
\begin{equation} 
\label{alkoferraya} 
\frac{1}{4\pi}{\cal G}(x) := \alpha(x) = \frac{a_0 + a_1 x}{1 + a_2 x + a_3 
x^2 + a_4 x^3} 
+ \frac{\pi \, \gamma_m}{\ln(e+x)}\,, 
\end{equation} 
\begin{equation} 
\label{arparams} 
\begin{array}{ccccc} 
a_0 & a_1 & a_2 & a_3 & a_4\\\hline 
1.47\;\; & 0.881 \;\; & 0.314 \;\; & 0.00986 \;\; & 0.00168 
\end{array}\,. 
\end{equation} 
Aspects of these DSE studies are in qualitative agreement with quenched
lattice-QCD results \cite{latticegluon}; e.g., the feature that the
dressed-gluon propagator is finite at $Q^2=0$. This effective interaction
yields the results in the second row of Table \ref{tab2}: apparently, $f_\pi$
is too small by a factor of $\sim 2$ and the scale of DCSB too small by a
factor of $\sim 4 =1.6^3$ \cite{pctprivate}.  This outcome could have been
anticipated from Ref.\ \cite{hmr98}.
 
Equations (\ref{dressedL}), (\ref{rainbow}) are correct on the ultraviolet
domain in which perturbation theory is applicable.  However, modelling is
involved in extrapolating onto the infrared domain.  One may require that
multiplicative renormalisability of the dressed fermion propagator be
preserved in this modelling \cite{cp90}.  The {\it Ansatz} of Ref.\
\cite{bloch} incorporates this constraint, however, it implicitly specifies a
kernel in the gap equation that violates a more important constraint, namely,
Lorentz covariance \cite{burden93}.  A minimal {\it Ansatz} that satisfies
both constraints and Eq.\ (\ref{dressedL}) finds expression in the kernel of
the renormalised gap equation as
\begin{equation} 
4\pi \alpha(Q^2)\frac{1}{Z^2(Q^2)} \,D_{\mu\nu}^{\rm free}(Q) 
 \frac{\lambda^a}{2}\gamma_\mu S(q) 
\frac{\lambda^a}{2}\gamma_\nu\,. \label{rainbow2} 
\end{equation} 
The gap equation's solution now gives the results in the third row of Table
\ref{tab2}.  [We simplified the numerical study by using a well-tried angle
approximation: $\alpha((k-q)^2)/ Z^2((k-q)^2)$ $\approx$
\mbox{$\alpha((k-q)^2)/ Z^{2}({\rm max}(k^2,q^2))$}.]
This procedure is evi\-dent\-ly a candidate for providing a bridge between
DSE studies of the gauge sector and what is known, from the calculation of
observables, to be the infrared strength required in the effective
interaction.  The increased interaction tension arises because the
enhancement of $1/Z(p^2)$, evident in Fig.\ \ref{figZ}, persists with this
kernel and amplifies the effective coupling: the interaction tension
calculated with Eq.\ (\ref{alkoferraya}) alone is $\sigma^\Delta =12\,$GeV/fm
but using Eq.\ (\ref{rainbow2}) we estimate that the $1/Z^2(Q^2)$ factor
raises this to $\sigma^\Delta \sim 25\,$GeV/fm.
 
\section{Confinement} 
This is a statement about the properties of coloured Sch\-win\-ger functions;
i.e., gauge-dependent quantities.  Hence it is plausible that the expression
of confinement will depend on the gauge fixing procedure one employs.  Linear
covariant gauges are most widely employed in DSE studies.  While the problem
of Gribov copies is unresolved, it affects only the far infrared behaviour of
coloured Schwinger functions and hence is hoped to have little impact on
physical observables. This conjecture is supported by the close
correspondence between Landau and Laplacian gauge two-point quark and gluon
Schwinger functions obtained in quenched lattice-QCD simulations
\cite{tonylaplacian}.  (Laplacian gauge fixing is a nonlocal prescription
that is identical to Landau gauge at ultraviolet momenta. It is free of
Gribov copies.)
 
It has long been known \cite{gastao} that for confinement it is sufficient
that no coloured Schwinger function possess a spectral representation.  This
is equivalent to the statement that all coloured Schwinger functions violate
reflection positivity.  That property can manifest itself \textit{almost} as
simply as, say, dressed-gluons being described by a propagator with a large
width.  The connection between confinement and the violation of reflection
positivity has cleanly been illustrated in QED3 \cite{pieterQED3}, and it is
a result of DSE studies that a sufficiently large interaction tension will
always produce a dressed-quark propagator that violates reflection
positivity. (NB.\ Gap equation solutions that violate reflection positivity
also exhibit DCSB.  The converse is not necessarily true.)  Thus it is a
widespread conjecture that confinement in QCD is realised this way.
 
A Schwinger function that violates reflection positivity is easily
identified: it is sufficient that the function exhibit a maximum or
inflection at some spacelike four momentum-squared.  In lattice-QCD
simulations, this feature is evident in the two-point gluon Schwinger
function \cite{latticegluon} and incipient in the scalar functions
characterising the dressed-quark two-point function \cite{latticequark}.  It
is plain in our results for the dressed-quark propagator, Figs.\ \ref{figZ},
\ref{figM}, and apparent in the DSE result for the gluon two-point function
\cite{alkofer2}.  The DSE solution for the ghost two-point function also
violates reflection positivity but in this case it is manifest in the
strength of its singularity at $k^2=0$.
 
A consistent picture may be emerging.  It is complemented by the fact that
the colour-singlet $S$-matrix elements which describe physical processes do
not exhibit unphysical quark and gluon production thresholds when calculated
using coloured Schwinger functions that violate reflection positivity.  The
ideas we have described also ensure that coloured composites such as diquarks
do not appear in the spectrum \cite{truncscheme,will}.
 
\section{Epilogue} 
The momentum-dependent dressing of quark and gluon propagators is a fact.  It 
is certainly the keystone of DCSB and quite likely plays a central role in 
confinement. Its incorporation into models of low-energy phenomena facilitates 
the development of a qualitatively reliable understanding and makes possible an 
excellent description of experiment.  That has provided guidance for modelling 
the long-range behaviour of the interaction between light-quarks.  A remaining 
challenge is to calculate it and pro\-gress is being made. 
 
\medskip

\subsection*{Acknowledgments} 
%
We thank P.O.\ Bowman for the lattice data and P.C.\ Tandy for a careful
reading of the manuscript.  PM and AR are grateful for the hospitality and
support of the Phys\-ics Division at ANL
during visits in which part of this work was conducted.  This work was
supported by: Deutsche Forschungsgemeinschaft, contract no.\ Ro 1146/3-1; the
US Department of Energy (DOE), Nuclear Physics Division, contract no.\
W-31-109-ENG-38; DOE grant nos.\ DE-FG02-96-ER-40947, DE-FG02-97-ER-41048;
and benefited from the resources of the US National Energy Research
Scientific Computing Center.
%
%

\end{document}